\documentclass[preprint,aps,prd,showpacs,nofootinbib]{revtex4}
\usepackage{mathrsfs}
\usepackage{amsmath}
\usepackage{graphicx}
\usepackage{subfigure}

\newcommand{\bea}{\begin{eqnarray}}
\newcommand{\eea}{\end{eqnarray}}
\newcommand{\beq}{\begin{equation}}
\newcommand{\eeq}{\end{equation}}

\def\k{{\vec k}}
\def\x{{\vec x}}
\def\/{\over}

\begin{document}

\title{Spontaneous excitation of a uniformly accelerated atom coupled with vacuum Dirac field fluctuations }
\author{  Wenting Zhou$^{1}$ and Hongwei Yu$^{1,2,}$\footnote{Corresponding author}  }
\affiliation{$^1$ Department of Physics and Key Laboratory of Low Dimensional Quantum
Structures and Quantum Control of Ministry of Education,\\
Hunan Normal University, Changsha, Hunan 410081, China \\
$^2$ Center for Nonlinear Science and Department of Physics, Ningbo
University,  Ningbo, Zhejiang 315211, China}

\begin{abstract}

We study the spontaneous excitation of a uniformly accelerated two-level atom non-linearly coupled to vacuum Dirac field fluctuations using the formalism proposed by Dalibard, Dupont-Roc and Cohen-Tannoudji (DDC) and generalized by us to the present case in the current paper. We find that a cross term  involving both vacuum fluctuations and radiation reaction appears, which is absent  in the linear coupling cases such as an atom interacting with vacuum scalar or electromagnetic fluctuations. Furthermore, the contribution of this term actually dominates over that of radiation reaction. Thus, the mean rate of change of the atomic energy can no longer be distinctively separated into only the contributions of vacuum fluctuations and radiation reaction as in the scalar and electromagnetic cases where the coupling is linear. Our result shows that a uniformly accelerated atom interacting with vacuum Dirac fluctuations would spontaneously excite and a unique feature in sharp contrast to the scalar and electromagnetic cases is the appearance of a term in the excitation rate which is proportional to the quartic acceleration.
\end{abstract}
\maketitle

\baselineskip=16pt

\section{Introduction}

Spontaneous excitation is one of the most interesting and prominent phenomena in the interactions of atoms with radiation and so far mechanisms such as vacuum fluctuations \cite{Welton48,Compagno83}, radiation reaction \cite{Ackerhalt73} or a combination of them \cite{Milonni75,Milonni88}
have been proposed as its possible physical explanation.
The ambiguity arises as a result of the fact that there exists a freedom in the choice of the ordering of the atom and field variables in a Heisenberg approach to the problem. The controversy was resolved by Dalibard, Dupont-Roc and Cohen-Tannoudji (DDC)~\cite{Dalibard82,Dalibard84} who showed that when a symmetric operator ordering is chosen then the contributions of vacuum fluctuations and radiation reaction to an atomic observable can be distinctively separated when a linear coupling between the atom and the field is considered, and furthermore this separation makes them  separately Hermitian. Using the DDC prescription, one can show that for inertial ground-state atoms, the contributions of vacuum fluctuations and radiation reaction to the rate of change of the mean excitation energy cancel exactly and this cancellation forbids any transitions from the ground state and thus ensures atom's stability in vacuum. While for any initial excited state, the rate of change of atomic energy acquires equal contributions from vacuum fluctuations and from radiation reaction \cite{Audretsch94}. The DDC formalism was then fruitfully generalized, in the case of linear coupling, to study the radiative properties of atoms in non-inertial motion \cite{Audretsch951,Audretsch952,Passante98,Zhu06,Yu06,Zhu07,Rizzuto07,Rizzuto091,Rizzuto092,Zhu10,Rizzuto011} or in a thermal bath  \cite{Tomazelli,Yu-thermal} in a flat spacetime and static ones in a gravitational field which interact with vacuum quantum scalar and electromagnetic fields
\cite{Yu07-12,ds}. These studies have shed some light on our understanding of the Unruh effect and Hawking radiation from a different physical perspective.

In this paper, we plan to study the spontaneous excitation of a uniformly accelerated atom interacting nonlinearly with fluctuating vacuum Dirac fields. Our interest in this issue is two fold. First, in the case of an atom in interaction with Dirac fields, the simplest Lorentz scalar
interaction Hamiltonian that we can introduce is non-linear (as that will be given later),
and we want to see what happens to DDC formalism when linear couplings are replaced by a non-linear
one, e.g, will the contribution to the rate of change of the mean atomic energy still be distinctively
separated into vacuum fluctuations and radiation reaction?
Second, when we go from the scalar field to the electromagnetic field, we get extra contribution proportional to the acceleration squared besides the thermal term characterized by the Plackian factor~\cite{Audretsch94,Audretsch951,Passante98,Zhu06}. It is then natural to wonder whether new features will appear for a spin $1/2$ Dirac field?

The paper is organized as follows. In Sec. II, we generalize the DDC formalism to Dirac fields where the coupling between an atom and the field is non-linear and then, in Sec. III, we use the generalized DDC formalism to calculate the spontaneous excitation rate of a uniformly accelerated atom with fluctuating Dirac fields in vacuum. We conclude in Sec. IV.  Natural units $\hbar=c=1$ and with metric signature $(+,-,-,-)$ will be used throughout the paper.

\section{General formalism}

Consider a two-level atom in interaction with vacuum Dirac field fluctuations in four dimensional Minkowski spacetime. Let $|-\rangle$, $|+\rangle$ denote the atomic ground state and excited one with energies being $-{1\/2}\omega_0$ and $+{1\/2}\omega_0$ respectively and $x^{\mu}=(x^0,\x)=(t,x,y,z)$ denote the Minkowski coordinates referring to an inertial reference frame. Assume that the atom is on a stationary trajectory $x(\tau)=(t(\tau),\x(\tau))$ where $\tau$ represents the proper time.

In the Dicke's notation \cite{Dicke}, the Hamiltonian that governs the evolution of the atom with respect to $\tau$ is
 \beq
H_A(\tau)=\omega_0R_3(\tau)\label{atomic Hamiltonian}
 \eeq
where
$R_3(0)=\frac{1}{2}|+\rangle\langle+|-\frac{1}{2}|-\rangle\langle-|$. The field $\psi(x)$ that the atom is assumed to be coupled to  satisfies the Dirac equation
 \beq
(i /\kern-0.50em \partial_{\mu}-m)\psi(x)=0
 \eeq
where $/\kern-0.50em \partial_{\mu}=\gamma^{\mu}\partial_{\mu}$ with $\partial_{\mu}={\partial\/\partial x^{\mu}}$ and
 \bea
\gamma^0=\begin{pmatrix}
  I & 0 \\
  0 & -I \\
\end{pmatrix}\;,\quad
\gamma^i=\begin{pmatrix}
  0 & \mathbf{\sigma}_i \\
  -\mathbf{\sigma}_i & 0 \\
\end{pmatrix}
 \eea
with $\mathbf{\sigma}$ being the Pauli matrices. The $\gamma$ matrices satisfy the algebra: $\{\gamma^{\mu},\gamma^{\nu}\}=2g^{\mu\nu}$ where $\{\;,\;\}$ denotes the anti-commutator. Solving the Dirac equation,  we can expand the field operator
in terms of a complete set of plane-wave solutions as
  \beq
\psi(x)=\sum_s\int {d^3\k\/(2\pi)^{3/2}}\sqrt{{m\/\omega_{\k}}}\;
           [b(\k,s,t)u(\k,s)e^{i\k\cdot\x}+d^\dagger(\k,s,t)v(\k,s)e^{-i\k\cdot\x}]
  \eeq
in which $b(\k,s,t)$ and $d^+(\k,s,t)$ are respectively annihilation and creation operators of particles and antiparticles with momentum $\vec{k}$ and spin $s$ and
 \bea
u(\k,s)&=&{{/\kern-0.50em k+m}\/\sqrt{2m(\omega_{\k}+m)}}u(0,s)\;,\\
v(\k,s)&=&{{-/\kern-0.50em k+m}\/\sqrt{2m(\omega_{\k}+m)}}v(0,s)\;
 \eea
where $/\kern-0.50em k=k^{\mu}\gamma_{\mu}$, $u(0,s)$ and $v(0,s)$ are the unit spinors in the particle rest frame with $k=(m,\mathbf{0})$ which have only upper or only lower components. When there is no coupling between the atom and the field, $b(\k,s,t)=b(\k,s)e^{-i\omega t}, d(\k,s,t)=d(\k,s)e^{-i\omega t}$. The vacuum is then defined by the annihilation operators as
 \beq
b(\k,s)|0\rangle=d(\k,s)|0\rangle=0\;.
 \eeq
The annihilation and creation operators satisfy the anti-commuting relations with
 \bea
\{b(\k,s),b^+(\k',s')\}=\{d(\k,s),d^+(\k',s')\}=\delta^3(\k-\k')\delta_{ss'}
 \eea
and others being zero.

The free Hamiltonian that governs the evolution of the free Dirac field with respect to the proper time $\tau$ is given by
 \beq
H_F(\tau)=\sum_s\int
   d^3\k\;\omega_{\vec{k}}\;[b^+(\k,s)\;b(\k,s)+d^+(\k,s)\;d(\k,s)]{dt\/d\tau}\;.
 \eeq
The interaction Hamiltonian that describes the coupling between the atom and the field is assumed to be
 \beq
H_I(\tau)=\mu R_2(\tau)\bar{\psi}(x(\tau))\psi(x(\tau))
 \eeq
where $\bar{\psi}(x(\tau))=\psi^\dagger(x(\tau))\gamma^0$, $\mu$ is the coupling constant that is assumed to be small and $R_2(0)=\frac{1}{2}i[R_-(0)-R_+(0)]$ with $R_+(0)=|+\rangle\langle-|$ and $R_-(0)=|-\rangle\langle+|$ being the atomic raising and lowering operators respectively. These operators obey the angular momentum algebra, i.e., $[R_3,R_\pm]=\pm R_\pm$, $[R_+,R_-]=2R_3$ where $[\;,\;]$ denotes the commutator. Notice that here the Hamiltonian is quadratic in the field operator, so the coupling is non-linear in contrast to the case of scalar and electromagnetic fields~\cite{Audretsch94,Passante98} where the coupling is linear. With this kind of nonlinear interaction, atomic transitions can occur both via absorption and emission of Dirac particle-antiparticle pairs  and inelastic scattering of a particle or antiparticle even at the lowest order of perturbation. This is in contrast to the linear coupling case, e.g., a scalar field, where the quantum is singly absorbed or emitted and inelastic scattering occurs
only at higher orders. Let us note that this interaction Hamiltonian has been used for the Unruh particle detector which is aimed to
detect vacuum Dirac field fluctuations and responses of the detector are found~\cite{Iyer,Takggi,Langlois06}.

The total Hamiltonian of the system (atom+field) is composed of the above three parts
 \beq
H(\tau)=H_A(\tau)+H_F(\tau)+H_I(\tau)\;.
 \eeq
Starting from the above Hamiltonian, we can obtain the Heisenberg equations of motion for the dynamical variables of the atom and the field:
 \bea
{d R_{\pm}(\tau)\/d\tau}&=&\pm i\omega_0R_{\pm}(\tau)
                          +i\mu[R_2(\tau),R_{\pm}(\tau)]\bar{\psi}(x(\tau))\psi(x(\tau))\;,
                          \label{R diffeq}\\
{d R_{3}(\tau)\/d\tau}&=&i\mu[R_2(\tau),R_{3}(\tau)]\bar{\psi}(x(\tau))\psi(x(\tau))\;,\\
{d\;b(\k,s,t(\tau))\/dt}&=&-i\omega_{\k}b(\k,s,t(\tau))+i\mu R_2(\tau)[\bar{\psi}(x(\tau))
                          \psi(x(\tau)),b(\k,s,t(\tau))]{d\tau\/dt}\label{b diffeq}\;.
 \eea
The Heisenberg equation of motion of the annihilation operator of antiparticles is obtained by replacing $b(\k,s,t(\tau))$ with $d(\k,s,t(\tau))$ in Eq.~(\ref{b diffeq}). In the above equations, we have kept the commutators unevaluated for latter use.

The solutions of the above equations of motion can be split into two parts: the free part that exists even when there is no coupling between the atom and the field and the source part that induced by the interaction between them and characterized by the coupling constant $\mu$:
 \bea
R_{\pm}(\tau)&=&R^f_{\pm}(\tau)+R^s_{\pm}(\tau)\;,\nonumber\\
R_{3}(\tau)&=&R^f_{3}(\tau)+R^s_{3}(\tau)\;,\nonumber\\
b(\k,s,t(\tau))&=&b^f(\k,s,t(\tau))+b^s(\k,s,t(\tau))\;.\nonumber
 \eea
Then formal integration of Eqs.~(\ref{R diffeq})-(\ref{b diffeq}) gives
 \bea
R^f_{\pm}(\tau)&=&R^f_{\pm}(\tau_0)e^{\pm i\omega_0(\tau-\tau_0)}\;,\label{Rpmf}\\
R^s_{\pm}(\tau)&=&i\mu\int^{\tau}_{\tau_0}d\tau'\;[R^f_2(\tau'),R^f_{\pm}(\tau)]
\bar{\psi}^f(x(\tau'))\psi^f(x(\tau'))\;,\\
R^f_{3}(\tau)&=&R^f_{3}(\tau_0)\;,\\
R^s_{3}(\tau)&=&i\mu\int^{\tau}_{\tau_0}d\tau'[R^f_2(\tau'),R^f_{3}(\tau)]
\bar{\psi}^f(x(\tau'))\psi^f(x(\tau'))\;,\label{R3s}\\
b^f(\k,s,t(\tau))&=&b^f(\k,s,t(\tau_0))e^{-i\omega_{\k}(t(\tau)-t(\tau_0))}\;,\\
b^s(\k,s,t(\tau))&=&i\mu\int^{\tau}_{\tau_0}d\tau'
R^f_2(\tau')[\bar{\psi}^f(x(\tau'))\psi^f(x(\tau')),b^f(\k,s,t(\tau))]\;.
 \eea
In the source parts of the above solutions, all operators on the right-hand side have been replaced by their free parts as we have made them only accurate to the first order in the coupling constant $\mu$. Similarly, we can derive the corresponding free part and source part of the annihilation operators of antiparticles. So, the field operator can be expressed as,
 \beq
\psi(x(\tau))=\psi^f(x(\tau))+\psi^s(x(\tau))
\label{decompose}
 \eeq
with
 \bea
\psi^f(x(\tau))&=&\sum_s\int{d^3\k\/(2\pi)^{3/2}}\sqrt{{m\/\omega_{\k}}}\nonumber\\
           &&\;\;\times[b^f(\k,s)u(\k,s)e^{-i\omega_{\k}(t(\tau)-t(\tau_0))+i\k\cdot(\x(\tau)-\x(\tau_0))}\nonumber\\
           &&\;\;\;+d^{f\dagger}(\k,s)v(\k,s)e^{i\omega_{\k}(t(\tau)-t(\tau_0))-i\k\cdot(\x(\tau)-\x(\tau_0))}]\;,
           \label{psif}\\
\psi^s(x(\tau))&=&i\mu\int^{\tau}_{\tau_0}d\tau'R^f_2(\tau')
                  [\bar{\psi}^f(x(\tau'))\psi^f(x(\tau')),\psi^f(x(\tau))]\label{psis}\;.
 \eea

Our aim now is to study the roles played by vacuum fluctuations associated with $\psi^f(x(\tau))$ and radiation reaction associated with $\psi^s(x(\tau))$ in the evolution of the atom. So, we need to plug the decomposition (\ref{decompose}) into the Heisenberg equation of motion of the atomic Hamiltonian
 \beq
{dH_A(\tau)\/d\tau}=i\mu\omega_0[R_2(\tau),R_{3}(\tau)]\bar{\psi}(x(\tau))\psi(x(\tau))\;.
 \eeq
In so doing, an issue of operator ordering arises as the free part ${\psi}^f(x(\tau))$ and source part ${\psi}^s(x(\tau))$ no longer separately commute with the atomic operators.  Actually, we can write
 \beq
{dH_A(\tau)\/d\tau}=\biggl({dH_A(\tau)\/d\tau}\biggr)_{vf}+\biggl({dH_A(\tau)\/d\tau}\biggr)_{cross}
+\biggl({dH_A(\tau)\/d\tau}\biggr)_{rr}\label{diff atom energy}\;,
 \eeq
where
 \bea
\biggl({dH_A(\tau)\/d\tau}\biggr)_{vf}&=&i\mu\omega_0(\lambda\bar{\psi}^f(x(\tau))\psi^f(x(\tau))[R_2(\tau),R_{3}(\tau)]\nonumber\\
                    &&\quad\;\;+(1-\lambda)[R_2(\tau),R_{3}(\tau)]\bar{\psi}^f(x(\tau))\psi^f(x(\tau)))\;,\\
\biggl({dH_A(\tau)\/d\tau}\biggr)_{cross}&=&i\mu\omega_0(\lambda\bar{\psi}^f(x(\tau))\psi^s(x(\tau))[R_2(\tau),R_{3}(\tau)]\nonumber\\
                    &&\quad\;\;+(1-\lambda)[R_2(\tau),R_{3}(\tau)]\bar{\psi}^f(x(\tau))\psi^s(x(\tau))\nonumber\\
                    &&\quad\;\;+\lambda\bar{\psi}^s(x(\tau))\psi^f(x(\tau))[R_2(\tau),R_{3}(\tau)]\nonumber\\
                    &&\quad\;\;+(1-\lambda)[R_2(\tau),R_{3}(\tau)]\bar{\psi}^s(x(\tau))\psi^f(x(\tau)))
 \eea
and
 \bea
\biggl({dH_A(\tau)\/d\tau}\biggr)_{rr}&=&i\mu\omega_0(\lambda\bar{\psi}^s(x(\tau))\psi^s(x(\tau))[R_2(\tau),R_{3}(\tau)]\nonumber\\
                    &&\quad\;\;+(1-\lambda)[R_2(\tau),R_{3}(\tau)]\bar{\psi}^s(x(\tau))\psi^s(x(\tau)))\;.
 \eea
Classically, for any $\lambda$ ranging from zero to one, the above equations are equivalent, but quantum mechanically they are not. We will chose $\lambda=1/2$, since such a choice ensures that the above three parts are all separately Hermitian. A few comments are now in order. First, the rhs of the first and the last equations only involve the free part and source part respectively and so
they can be interpreted as the sole contribution of vacuum fluctuations and that of the radiation reaction respectively. Second, even if we chose a symmetric ordering, there still exists a contribution to the rate of change of the atomic energy that involves both the free part and source part and this cross term can neither be understood as the sole contribution of vacuum fluctuations nor that of the radiation reaction, but rather a combination of them. This is in sharp contrast to the cases of linear couplings where no cross term appears and the contribution to the rate of change of the atomic energy can be distinctively separated into only vacuum fluctuations and radiation reaction \cite{Audretsch94,Zhu06}. Furthermore, as we will see later, the contribution of this cross term dominates over that of radiation reaction. In fact, the appearance of a cross-terms is a direct result of the nonlinear coupling between the atom and the field.

Taking the average value of the rate of change of the atomic energy, Eq.~(\ref{diff atom energy}), over the  state of the system, $|0,b\rangle$, where $0$ represents the vacuum state of the field and $b$ the state of the atom, we find that both $({dH_A(\tau)\/d\tau})_{vf}$ and $({dH_A(\tau)\/d\tau})_{cross}$ are of the order of $\mu^2$, whereas $({dH_A(\tau)\/d\tau})_{rr}$ is of $\mu^3$. So, to the order $\mu^2$, we have
 \beq
\biggl\langle{dH_A(\tau)\/d\tau}\biggr\rangle=\biggl\langle{dH_A(\tau)\/d\tau}\biggr\rangle_{vf}
                                               +\biggl\langle{dH_A(\tau)\/d\tau}\biggr\rangle_{cross}
 \eeq
with
 \bea
\biggl\langle{dH_A(\tau)\/d\tau}\biggr\rangle_{vf}
&=&{1\/2}i\mu\omega_0\langle\bar{\psi}^f(x(\tau))\psi^f(x(\tau))[R_2(\tau),R_{3}(\tau)]
+[R_2(\tau),R_{3}(\tau)]\bar{\psi}^f(x(\tau))\psi^f(x(\tau))\rangle\;,\nonumber\\ \\
\biggl\langle{dH_A(\tau)\/d\tau}\biggr\rangle_{cross}
&=&{1\/2}i\mu\omega_0\langle\bar{\psi}^f(x(\tau))\psi^s(x(\tau))[R_2(\tau),R_{3}(\tau)]
+[R_2(\tau),R_{3}(\tau)]\bar{\psi}^f(x(\tau))\psi^s(x(\tau))\nonumber\\
&&\quad\quad+\bar{\psi}^s(x(\tau))\psi^f(x(\tau))[R_2(\tau),R_{3}(\tau)]
+[R_2(\tau),R_{3}(\tau)]\bar{\psi}^s(x(\tau))\psi^f(x(\tau))\rangle\;.\nonumber\\
 \eea
Here $\langle\;\rangle$ represents the expectation value over the state of the system. A distinct feature as compared to the scalar and electromagnetic cases where the couplings are linear \cite{Audretsch94,Zhu06} is that now the contribution of radiation reaction is of the order higher than that of the vacuum fluctuations and thus negligible. Further simplifications by using Eqs.~(\ref{Rpmf})-(\ref{R3s}), (\ref{psif}) and (\ref{psis}) yield
 \bea
\biggl\langle{dH_A(\tau)\/d\tau}\biggr\rangle_{vf}&
=&2i\mu^2\int^{\tau}_{\tau_0}d\tau'\;C^F(x(\tau),x(\tau')){d\/d\tau}\chi^A(\tau,\tau')\;,
\label{general vf contribution}\\
\biggl\langle{dH_A(\tau)\/d\tau}\biggr\rangle_{cross}&
=&2i\mu^2\int^{\tau}_{\tau_0}d\tau'\;\chi^F(x(\tau),x(\tau')){d\/d\tau}C^A(\tau,\tau')
\label{general rr contribution}
 \eea
with
 \bea
C^F(x(\tau),x(\tau'))&=&\frac{1}{2}\langle0|\{\bar{\psi}^f(x(\tau))
                        \psi^f(x(\tau)),\bar{\psi}^f(x(\tau'))\psi^f(x(\tau'))\}
|0\rangle\;,\\
\chi^F(x(\tau),x(\tau'))&=&-\frac{1}{2}\langle0|(\bar{\psi}^f(x(\tau))
                           [\bar{\psi}^f(x(\tau'))\psi^f(x(\tau')),\psi^f(x(\tau))]\nonumber\\
                           &&\quad\;\;\;\;+[\bar{\psi}^f(x(\tau'))\psi^f(x(\tau')),\bar{\psi}^f(x(\tau))]
                           \psi^f(x(\tau)))|0\rangle
 \eea
being the two statistical functions of the field and
 \bea
C^A(\tau,\tau')&=&\frac{1}{2}\langle
b|\{R_2^f(\tau),R_2^f(\tau')\}|b\rangle\;,\\
\chi^A(\tau,\tau')&=&\frac{1}{2}\langle
b|[R_2^f(\tau),R_2^f(\tau')]|b\rangle
 \eea
being the two susceptibility functions of the atom which can be further simplified to be
 \bea
C^A(\tau,\tau')&=&\frac{1}{2}\sum_{d}|\langle
b|R_2(0)|d\rangle|^2\,
   (e^{i\omega_{bd}(\tau-\tau')}+e^{-i\omega_{bd}(\tau-\tau')})\;,\label{ca}\\
\chi^A(\tau,\tau')&=&\frac{1}{2}\sum_{d}|\langle
b|R_2(0)|d\rangle|^2\,
   (e^{i\omega_{bd}(\tau-\tau')}-e^{-i\omega_{bd}(\tau-\tau')})\;.\label{chia}
 \eea
The summation in the above two equations extends over the complete set of the states of the atom.

\section{Spontaneous excitation of a uniformly accelerated atom}

Assume that a two-level atom is initially in state $|b\rangle$ and is uniformly accelerated along the stationary trajectory
 \beq
t(\tau)=\frac{1}{a}\sinh(a\tau)\;,\quad\;x(\tau)
=\frac{1}{a}\cosh(a\tau)\;,\quad\;y(\tau)=z(\tau)=0
 \eeq
with acceleration $a$ in four dimensional Minkowski vacuum with quantum Dirac field fluctuations.
An observer moving along such a trajectory is usually called the Rindler observer.

To evaluate the rate of change of the atomic energy by the formalism generalized in the preceding section, we should first calculate  two statistical functions of the Dirac field. For this purpose, let us introduce the following matrix with respect to the two point function of the Dirac field as it would be useful in latter calculations
 \bea
S^+_n(x(\tau),x(\tau'))&=&\langle0|\psi(x(\tau))\bar{\psi}(x(\tau'))|0\rangle\nonumber\\
                       &=&\int{d^3\k\/(2\pi)^3}{m\/\omega_{\k}}\sum_su(\k,s)\bar{u}(\k,s)\;
                       e^{-i\omega_{\k}(t(\tau)-t(\tau'))+i\k\cdot(\x(\tau)-\x(\tau'))}\;.
 \eea
Generically, the above matrix is not stationary even if the world line was. The reason is as follows.  the spinor, $\psi(x(\tau))$, is not
\textquotedblleft intrinsic\textquotedblright to the observer moving along the world line. It will in general
\textquotedblleft rotate\textquotedblright with respect to the observer's proper reference frame~\cite{Takggi,Misner}, i.e., the reference frame Fermi-Walker transported by the observer. To keep the spinor, as it were, \textquotedblleft in the same direction\textquotedblright with respect to the proper frame reference, it is useful to introduce a transformation to the Dirac field operator~\cite{Misner,Gross}
 \bea
\psi(x(\tau))&\rightarrow& S_{\tau}\psi(x(\tau))\;,\\
\bar{\psi}(x(\tau))&\rightarrow&\bar{\psi}(x(\tau))S^{-1}_{\tau}\;,
 \eea
where the matrix $S_{\tau}$ is to take care of the Fermi-Walker transportation and it is given,  in the case of a uniformly accelerated worldline with proper acceleration $a$,  by
\beq
S_{\tau}=e^{{1\/2}a\tau\gamma_0\gamma_1}=\cosh({a\tau/2})+\gamma_0\gamma_1\sinh({a\tau/2})\;.
 \eeq
Thus it is more appropriate to define a new matrix \cite{Takggi,Misner}
 \bea
g(\tau,\tau')=S_{\tau}S^+_n(x(\tau),x(\tau'))S_{\tau'}^{-1}\label{g Delta}
 \eea
which will be later shown to be a function of the time interval $\Delta\tau=\tau-\tau'$ and can effectively simplify the calculations concerned. Similarly, we can define another matrix with respect to the two point function of the field as follows
 \bea
(S_n^-(x(\tau'),x(\tau)))_{ab}&=&\langle0|\bar{\psi}_b(x(\tau))\psi_a(x(\tau'))|0\rangle\nonumber\\
                       &=&\int{d^3\k\/(2\pi)^3}{m\/\omega_{\k}}\sum_s\bar{v}_b(\k,s)v_a(\k,s)
                       e^{-i\omega_{\k}(t(\tau)-t(\tau'))+i\k\cdot(\x(\tau)-\x(\tau'))}\;.
 \eea
Noticing the relations that
 \bea
\sum_su(\k,s)\bar{u}(\k,s)&=&{/\kern-0.50em k+m\/2m}\;,\\
\sum_sv(\k,s)\bar{v}(\k,s)&=&{/\kern-0.50em k-m\/2m}\;
 \eea
we find
 \bea
S_n^+(x(\tau),x(\tau'))&=&(i/\kern-0.50em \partial+m)G^{+}(x(\tau),x(\tau'))\;,\label{sn+}\\
S_n^-(x(\tau),x(\tau'))&=&-(i/\kern-0.50em \partial+m)G^{-}(x(\tau),x(\tau'))\label{sn-}\;,
 \eea
where
 \beq
G^{+}(x(\tau),x(\tau'))=\frac{1}{(2\pi)^3}\int{d^3\k\/2\omega_{\k}}\;
e^{-i\omega_{\k}(t(\tau)-t(\tau'))+i\k\cdot(\x(\tau)-\x(\tau'))}\label{G+}
 \eeq
and $G^{-}(x(\tau),x(\tau'))=G^{+}(x(\tau'),x(\tau))$ are just Wightman functions of the scalar field in four dimensional Minkowski spacetime. We can deduce from these relations that them atrices $S_n^+$ and $S_n^-$ are related as
 \beq
S_n^-(x(\tau),x(\tau'))=S_n^+(x(\tau'),x(\tau))|_{m\rightarrow -m}\;.
 \eeq
By using the properties $S_{\tau}^{-1}=S_{-\tau}$, $S_{\tau}S_{\tau'}=S_{\tau+\tau'}$, and $\gamma_0S_{\tau}=S_{-\tau}\gamma_0$, it can be proved that the function $g(\tau,\tau')$ can be expressed in terms of the interval of the atomic proper time, $\Delta\tau=\tau-\tau'$, as
 \bea
g(\tau,\tau')&=&g(\Delta\tau)\nonumber\\
             &=&(-\gamma^0\partial_z+mS_{\Delta\tau})\;G^+(z(\Delta\tau))\label{g}
 \eea
with
 \bea
G^+(z(\Delta\tau))&=&\frac{1}{(2\pi)^3}\int{d^3\k\/2\omega_{\k}}\;e^{-\omega_{\vec{k}}z(\Delta\tau)}\;,\\
z(\Delta\tau)&=&i{2\/a}\sinh\biggl({a\/2}\Delta\tau-i\epsilon\biggr)\;.
 \eea
For the case of massless Dirac fields,
 \beq
G^+(z(\Delta\tau))={1\/4\pi^2[z(\Delta\tau)]^2}\;.
 \eeq

We can show, after lengthy simplifications, that the following functions related with the two statistical functions of the Dirac field can be expressed in terms of the above two matrices as
 \bea
\langle0|\bar{\psi}(x(\tau))\psi(x(\tau))\bar{\psi}(x(\tau'))\psi(x(\tau'))|0\rangle
&=&Tr[S_n^+(x(\tau),x(\tau'))S_n^-(x(\tau'),x(\tau))]\;,\label{ep sy}\\
\langle0|[\bar{\psi}(x(\tau'))\psi(x(\tau')),\bar{\psi}(x(\tau))]\psi(x(\tau))|0\rangle
&=&Tr[S_n^-(x(\tau),x(\tau'))S_n^+(x(\tau'),x(\tau))]\nonumber\\
&+&Tr[S_n^-(x(\tau),x(\tau'))S_n^-(x(\tau'),x(\tau))]\;,\label{fa1}\\
\langle0|\bar{\psi}(x(\tau))[\bar{\psi}(x(\tau'))\psi(x(\tau')),\psi(x(\tau))]|0\rangle
&=&-Tr[S_n^+(x(\tau),x(\tau'))S_n^-(x(\tau'),x(\tau))]\nonumber\\
&&-Tr[S_n^-(x(\tau'),x(\tau))S_n^-(x(\tau),x(\tau'))]\label{fa2}
 \eea
where $Tr[\cdots]$ represents the trace of a matrix. In obtaining the above three equations, we have used $(\gamma_0S_{\tau})^2=1$ and $[A,BC]=\{A,B\}C-B\{A,C\}$. In Eq.~(\ref{ep sy}), we have discarded an infinite constant as it would have no contributions to the integrals in the following calculations. Considering massless Dirac field fluctuations ($m=0$), we find
 \beq
Tr[S_n^+(x(\tau),x(\tau'))S_n^-(x(\tau'),x(\tau))]=4[\partial_zG^+(z(\Delta\tau))]^2\;.
 \eeq
Similar simplifications can also be obtained for Eqs.~(\ref{fa1}) and (\ref{fa2}). Consequently, the two statistical functions of the Dirac field can be shown to be given by
 \bea
C^F(x(\tau),x(\tau'))&=&-{a^6\/128\pi^4}\biggl[{1\/\sinh^6({a\/2}\Delta\tau-i\epsilon)}
+{1\/\sinh^6({a\/2}\Delta\tau+i\epsilon)}\biggr]\;,\label{cf}\\
\chi^F(x(\tau),x(\tau'))&=&-{a^6\/128\pi^4}\biggl[{1\/\sinh^6({a\/2}\Delta\tau-i\epsilon)}
-{1\/\sinh^6({a\/2}\Delta\tau+i\epsilon)}\biggr]\;.\label{chif}
 \eea

Inserting Eqs.~(\ref{cf}) and (\ref{chia}) into Eq.~(\ref{general vf contribution}) and taking the proper time to be infinite long, we can express the contributions of vacuum fluctuations to the mean rate of change of the atomic energy as
 \bea
\biggl\langle{dH_A(\tau)\/d\tau}\biggr\rangle_{vf}&=&
{\mu^2a^6\/128\pi^4}\sum_{d}|\langle b|R_2(0)|d\rangle|^2\omega_{bd}
\nonumber\\&&\quad\quad\times
\int^{\infty}_{0}d\Delta\tau\biggl[{1\/\sinh^6({a\/2}\Delta\tau-i\epsilon)}
+{1\/\sinh^6({a\/2}\Delta\tau+i\epsilon)}\biggr]
(e^{i\omega_{bd}\Delta\tau}+e^{-i\omega_{bd}\Delta\tau})\;.
\nonumber\\
 \eea
By exploiting the techniques of contour integration and residue theory, the above integral can be calculated out to be
 \bea
\biggl\langle{dH_A(\tau)\/d\tau}\biggr\rangle_{vf}
=&-&{\mu^2\/120\pi^3}\sum_{\omega_b>\omega_d}|\langle b|R_2(0)|d\rangle|^2\omega_{bd}^6
\biggl(1+5{a^2\/\omega_{bd}^2}+4{a^4\/\omega_{bd}^4}\biggr)
\biggl(1+{2\/{e^{2\pi\omega_{bd}/a}-1}}\biggr)\nonumber\\
&+&{\mu^2\/120\pi^3}\sum_{\omega_b<\omega_d}|\langle b|R_2(0)|d\rangle|^2\omega_{bd}^6
\biggl(1+5{a^2\/\omega_{bd}^2}+4{a^4\/\omega_{bd}^4}\biggr)
\biggl(1+{2\/{e^{2\pi|\omega_{bd}|/a}-1}}\biggr)\;.
\nonumber\\ \label{final vf contribution}
 \eea
Obviously, just as that of a uniformly accelerated atom interacting with vacuum scalar or electromagnetic fluctuations (see Eq.~(56) in Ref.~\cite{Audretsch94} and Eq.~(28) in Ref.~\cite{Zhu06}), the contributions of vacuum Dirac fluctuations would raise the atomic energy if the atom is initially in its ground state (concerned with the term $(\omega_b<\omega_d)$) and diminish its energy if the initial state of the atom is the excited state (concerned with the term $(\omega_b>\omega_d)$). A sharp feature in contrast to both the cases of an atom in interaction with massless scalar field and electromagnetic field fluctuations is the appearance of a term proportional to $a^4$ which is unique to the Dirac fluctuations.


Similarly, we can find, by taking Eqs.~(\ref{chif}) and (\ref{ca}) into Eq.~(\ref{general rr contribution}), the contributions of the cross term to the mean rate of change of the atomic energy as
 \bea
\biggl\langle{dH_A(\tau)\/d\tau}\biggr\rangle_{cross}&=&
{\mu^2a^6\/128\pi^4}\sum_{d}|\langle b|R_2(0)|d\rangle|^2\omega_{bd}
\nonumber\\&&\quad\quad\times
\int^{\infty}_{0}d\Delta\tau\biggl[{1\/\sinh^6({a\/2}\Delta\tau-i\epsilon)}
-{1\/\sinh^6({a\/2}\Delta\tau+i\epsilon)}\biggr]
(e^{i\omega_{bd}\Delta\tau}-e^{-i\omega_{bd}\Delta\tau})\;.
\nonumber\\
 \eea
Further calculations give
 \bea
\biggl\langle{dH_A(\tau)\/d\tau}\biggr\rangle_{cross}&
=&-{\mu^2\/120\pi^3}\sum_{\omega_b<\omega_d}|\langle b|R_2(0)|d\rangle|^2\omega_{bd}^6
\biggl(1+5{a^2\/\omega_{bd}^2}+4{a^4\/\omega_{bd}^4}\biggr)\nonumber\\
&&-{\mu^2\/120\pi^3}\sum_{\omega_b>\omega_d}|\langle b|R_2(0)|d\rangle|^2\omega_{bd}^6
\biggl(1+5{a^2\/\omega_{bd}^2}+4{a^4\/\omega_{bd}^4}\biggr)\;.\label{final rr contribution}
 \eea
It is interesting to note that the contribution of the cross term here always diminishes the
atomic energy no matter if the initial state of the atom is the ground state or the excited state,
so the cross term just plays the role as that played by the radiation reaction in the scalar
and electromagnetic field cases \cite{Audretsch94,Zhu06}. The presence of it ensures that the
inertial ground state atoms will be stable in the Minkowski vacuum since now the contribution
of radiation reaction is of order higher than that of the vacuum fluctuations and thus negligible.

Adding up Eqs.~(\ref{final vf contribution}) and (\ref{final rr contribution}), we obtain the
total mean rate of change of the atomic energy
 \bea
\biggl\langle{dH_A(\tau)\/d\tau}\biggr\rangle_{tot}
=&-&{\mu^2\/60\pi^3}\sum_{\omega_b>\omega_d}|\langle b|R_2(0)|d\rangle|^2\omega_{bd}^6
\biggl(1+5{a^2\/\omega_{bd}^2}+4{a^4\/\omega_{bd}^4}\biggr)
\biggl(1+{1\/{e^{2\pi\omega_{bd}/a}-1}}\biggr)\nonumber\\
&+&{\mu^2\/60\pi^3}\sum_{\omega_b<\omega_d}|\langle b|R_2(0)|d\rangle|^2\omega_{bd}^6
\biggl(1+5{a^2\/\omega_{bd}^2}+4{a^4\/\omega_{bd}^4}\biggr)
{1\/{e^{2\pi|\omega_{bd}|/a}-1}}\;.
 \eea
The above result reveals that uniformly accelerated atoms in ground state in interaction with fluctuating Dirac fields in vacuum would spontaneously excite. This is consistent with the response of Dirac particle detector found in Ref.~\cite{Langlois06}. However, our result can be considered as providing a transparent underlying physical mechanism on why such a particle detector clicks. A distinct feature is the appearance of the $a^4$ term in addition to the Planckian thermal factor which can be viewed as a result of an ambient thermal bath at the Unruh temperature at $T=a/2\pi$. This term is absent in both the scalar and electromagnetic cases. For a typical transition frequency of a hydrogen atom, $\omega\sim10^{16}s^{-1}$, the corrections due to the acceleration are negligible when $a\ll\omega\sim10^{24}m/s^2$ and it becomes appreciable when the acceleration  approaches  $\sim10^{24}m/s^2$. In fact, the contribution of the $a^4$ term which is unique to the current case becomes dominant when $a\gg\omega$. Here, it is also interesting to note that the response of an accelerated detector to the Dirac vacuum fluctuations near a fluctuating event horizon has recently been studied~\cite{Svaiter12}.

 The above result reduces in the limit $a\rightarrow 0$ to
 \beq
\biggl\langle{dH_A(\tau)\/d\tau}\biggr\rangle_{tot}
=-{\mu^2\/60\pi^3}\sum_{\omega_b>\omega_d}|\langle b|R_2(0)|d\rangle|^2\omega_{bd}^6\;.
 \eeq
This means that no spontaneous excitation would occur for inertial atoms in their ground states
since the term associated with $\omega_b<\omega_d$ vanishes, as a result of the complete cancellation of the contribution of vacuum fluctuations and  that of the cross term. Without the cross term, inertial ground state atoms in vacuum would not be stable. This is in sharp contrast to the scalar and electromagnetic cases that it is the cancellation of contribution of radiation reaction which is negligible in the present case and that of the vacuum fluctuations that ensures the stability of the inertial ground state atoms in vacuum \cite{Audretsch94,Zhu06}.

\section{Summary}

We have generalized the DDC formalism to the case of an atom interacting with vacuum Dirac field fluctuations where the coupling between the atom and field is non-linear, and find that a cross term appears, which involves both vacuum fluctuations and radiation reaction and which  is absent in the linear coupling cases such as an atom interacting with vacuum scalar or electromagnetic fluctuations. Furthermore, the contribution of this term actually dominates over that of radiation reaction. Thus, the mean rate of change of the atomic energy can no longer be distinctively separated into only the contributions of vacuum fluctuations and radiation reaction as in the scalar and electromagnetic cases where the coupling is linear. But rather,  the evolution of the atom is governed in the leading order by  vacuum fluctuations and the cross term that is a combined effect of  vacuum fluctuations and radiation reaction.

We then calculated the spontaneous excitation of a uniformly accelerated atom in interaction  with fluctuating vacuum Dirac fields.
Our result shows that such a uniformly accelerated atom would spontaneously excite in vacuum and the excitation rate contains, besides what can be viewed as a result of an ambient thermal bath at the Unruh temperature at $T=a/2\pi$,  a term proportional to quartic acceleration which is absent in both the scalar and electromagnetic cases.

\begin{acknowledgments}

This work was supported in part by the National Natural Science Foundation of China under Grants No. 11075083, and No. 10935013; the Zhejiang Provincial Natural Science Foundation of China under Grant No. Z6100077; the National Basic Research Program of China under Grant No. 2010CB832803; the PCSIRT under Grant No. IRT0964;  the Hunan Provincial Natural Science Foundation of China under Grant No. 11JJ7001; and Hunan Provincial Innovation Foundation For
Postgraduate under Grant No. CX2011B187.

\end{acknowledgments}

\end{document}